\documentclass[11pt]{article}
\usepackage{blois,epsfig}

\def\Journal#1#2#3#4{{#1} {\bf #2}, #3 (#4)}

\def\APP{\em Astropart. Phys.  }
\def\EPJC{{\em Eur. Phys. J.} C}
\def\JPG{{\em J. Phys.} G}
\def\JCAP{\em JCAP}

\def\NPB{{\em Nucl. Phys.} B}

\def\PRL{\em Phys. Rev. Lett.}
\def\PRD{{\em Phys. Rev.} D}

\def\be{\begin{equation}}
\def\ee{\end{equation}}
\def\bea{\begin{eqnarray}}
\def\eea{\end{eqnarray}}

\begin{document}
\vspace*{4cm}
\title{LATEST RESULTS FROM XENON100 DATA}

\author{ L. SCOTTO LAVINA \footnote{scotto@subatech.in2p3.fr} (for the XENON Collaboration)}

\address{Laboratoire SUBATECH, CNRS/In2p3, 44307 Nantes, France}

\maketitle\abstracts{
XENON100 is the current phase of the XENON dark matter program, which aims for the direct detection of WIMPs with liquid xenon time-projection chambers. We present the status of the experiment after 224.6 live days taken in 2011 and 2012 during which the detector successfully improved in terms of more calibration data, higher xenon purity, lower threshold and better background removal. The analysis has yielded no evidence for dark matter interactions. The status of the next generation XENON1T detector will be briefly described.}

\section{Introduction}
About eighty years since it was postulated, astronomical and cosmological observations indicate that a large amount of the content of the Universe is made of Dark Matter ~\cite{pdg}. While the presence of large quantities of Dark Matter is well established, its nature is still unknown. The most widely discussed models for non-baryonic dark matter are based on the Cold Dark Matter hypothesis, and the corresponding candidate particle arising in theories beyond the Standard Model has the generic name of Weakly Interacting Massive Particle (WIMP) ~\cite{wimp}. The search for these particles is performed with a wide variety of experimental approaches. In direct detection experiments, one attempts to observe the nuclear recoils produced by WIMPs scattering off nucleons ~\cite{wimpscattering}. The expected signal presents a recoil spectrum which falls exponentially with energy and extends to a few tens of keV only. The expected low event rate requires large detectors built from radio-pure materials and with good background rejection capabilities.

Liquid xenon (LXe) has ideal properties as a dark matter target and it is today used by many experiments to build massive, homogeneous and position-sensitive detectors. It has high scintillation and ionization yields because of its low ionization potential.

The XENON program is a phased approach to WIMP direct detection with time projection chambers (TPCs) using ultra-pure LXe as both target and detection medium by employing an increasing fiducial target mass scale (10kg, 100kg and 1000kg).
%To test the concept of the program, a detector with a fiducial mass on the order of 10 kg, XENON10 ~\cite{xenon10}, was developed and operated at the Gran Sasso National Laboratory (LNGS), in Italy.
The present phase is the XENON100 detector ~\cite{xenon100detector}, which has been taking data at the Gran Sasso National Laboratory (LNGS), in Italy, since 2008.
%Sections 2 and 3 describe the XENON100 experiment and its recent results. The next step is XENON1T ~\cite{xenon1t}, having a total LXe mass of 2.4 tons, which is already approved to be built in Hall B at LNGS and construction will start in late 2012. Section 4 draws a brief overview of the status of the XENON1T experiment.

\section{The XENON100 experiment}\label{subsec:xenon100}

The XENON100 detector is a cylindrical two-phase TPC enclosing a LXe target mass of 62 kg with an additional 99 kg, optically separated from the target, instrumented as an active scintillator veto. The TPC and the veto are mounted in a double-walled stainless-steel cryostat, enclosed by a passive shield made from OFHC copper, polyethylene, lead and water/polyethylene. The shield is continuously purged with boil-off $\rm N_2$ gas in order to suppress radon background. The LXe is kept at the operating temperature of about $-91\,^{\circ}{\rm C}$ by a pulse tube refrigerator mounted outside the shield. The detector details are described in Aprile {\it et al} ~\cite{xenon100detector}.

An interaction within the active volume of the detector creates ionization electrons and prompt scintillation photons. The electrons drift through the liquid under an external electric field with a speed of about 2 mm/$\mu$s. They then are accelerated by a stronger field and extracted into the gaseous phase above the liquid, where they generate secondary proportional scintillation light. Two arrays of photomultiplier tubes, one in the liquid and one in the gas, detect the prompt scintillation (S1) and the delayed secondary scintillation signal (S2). The time difference between these two signals gives the depth of the interaction in the TPC with a resolution of 0.3 mm (1$\sigma$). The hit pattern of the S2 signal on the top array allows reconstruction of the horizontal position of the interaction vertex with a resolution $<3$ mm (1$\sigma$). Therefore, the TPC yields a very good three-dimensional event localization, enabling rejection of the majority of the background via simple fiducial volume cuts, thanks to the self-shielding capabilities of LXe. The ratio of the two signals is different for nuclear recoils (NR, created by WIMP or neutron interactions) and electronic recoils (EL, produced by beta and gamma rays), providing the basis of one of the major background discrimination techniques. Rejection of double scattering events further enhances the discrimination of WIMP signal from background. The XENON Collaboration recently published a paper ~\cite{xenon100analysis} describing the general methods developed so far for the analysis of XENON100 data, focusing on the 100.9 live days science run of 2010 from which results on spin-independent elastic ~\cite{xenon100run8} and inelastic ~\cite{xenon100run8inelastic} WIMP-nucleon cross-sections have been obtained.

\subsection{Improvements with the new run}\label{subsec:xenon100latest}

The latest XENON100 run lasted 13 months from February 2011 to March 2012. Compared to the data from the previous run, the new dark matter search is characterized by the following improved experimental conditions ~\cite{xenon100run10}:

Larger exposure. Besides 3 interruptions due to equipment maintenance, the data were acquired continuously. Dark matter data taking was otherwise only interrupted by regular calibrations (using blue LED light, $^{137}$Cs, $^{60}$Co, $^{232}$Th and $^{241}$AmBe sources). Periods with increased electronic noise or very localized light emission in the xy-plane were removed from the data, as well as periods in which detector parameters such as temperature or pressure fluctuated outside of their normal range. This results in a final dark matter dataset of 224.6 live days.

Lower background. We obtained a substantial reduction of the intrinsic background from $^{85}$Kr by cryogenic distillation. The $^{nat}$Kr concentration in Xe has been lowered to (19$\pm$4)ppt, as measured in a Xe gas sample from the detector using ultra-sensitive rare gas mass spectrometry combined with a Kr/Xe separation technique.

Lower trigger. The data have been acquired under improved electronic noise conditions and with a modified hardware trigger to allow for a reduced S2 trigger threshold for $>$ 99\% of S2 signals above 150 photo-electrons (PE) to be recorded. This leads to virtually no loss of events in the energy region of interest.

\begin{figure}[t]
  \caption{Evolution of the electron lifetime during 2011-2012. Gray zones correspond to the maintenance periods
    \label{fig:electronlifetime}}
  \begin{center}
    \includegraphics[width=1.0\textwidth]{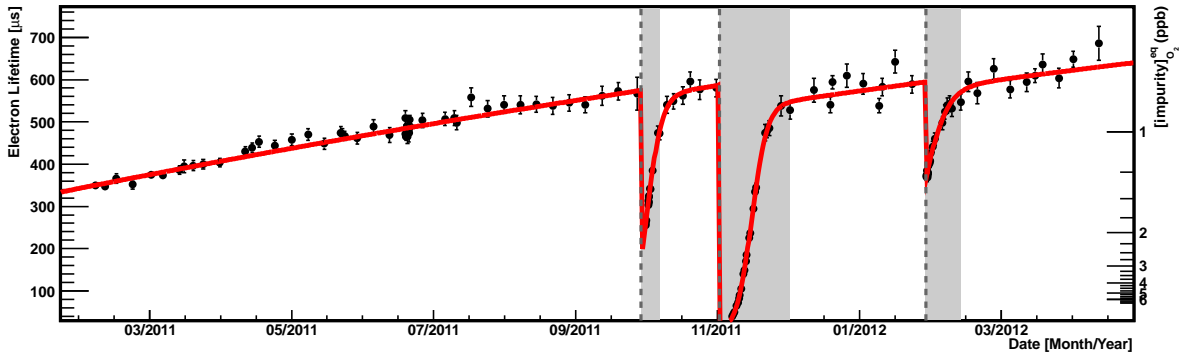}
  \end{center}
\end{figure}

Improved corrections on light collection. The non-uniform light collection by the two PMT arrays and the attenuation of the ionization signal by residual impurities lead to a position-dependent S1 and S2 signal response. The S1 and S2 yields are 3-dimensionally corrected using maps derived from calibration data in order to optimize the response very close to the PMTs. The electron lifetime $\tau_e$ ~\cite{xenon100analysis}, used to describe the ionization loss by impurities in LXe, was measured regularly with a $^{137}$Cs source throughout the data taking period. Figure \ref{fig:electronlifetime} shows an increase of $\tau_e$ along the whole run. The value increased from 374 $\mu s$ to 611 $\mu s$, with an average of $\tau_e$ = 514 $\mu s$, which corresponds to a level of electronegative impurities of about 1 ppb O$_2$-equivalent. Finally, the width of the S2 signal is also corrected such that it is independent of these inhomogeneities. The considerably larger amount of calibration data taken during this dark matter run allowed for an improvement of the accuracy in most of these corrections to the percent-level.

Improved background rejection cuts. In order to identify valid NR candidate events with the highest possible acceptance, several classes of cuts and event selections are applied to the data. The cuts and their acceptance determination are identical or similar to those used the previous runs ~\cite{xenon100analysis}, with the exception that the cut against electronic noise, which has been improved considerably. The fiducial volume used in this analysis contains 34 kg of LXe. The volume was determined by maximizing the sensitivity of this run given the accessible ER background above the blinding cut.

\subsection{Latest results}\label{subsec:xenon100latest}

\begin{figure}[t]
\label{fig:xenon100run10result}
\caption{(left) Space parameter indicating the search for WIMPs. (right) New result on spin-independent WIMP-nucleon scattering from XENON100}
\begin{center}
\includegraphics[width=0.45\textwidth]{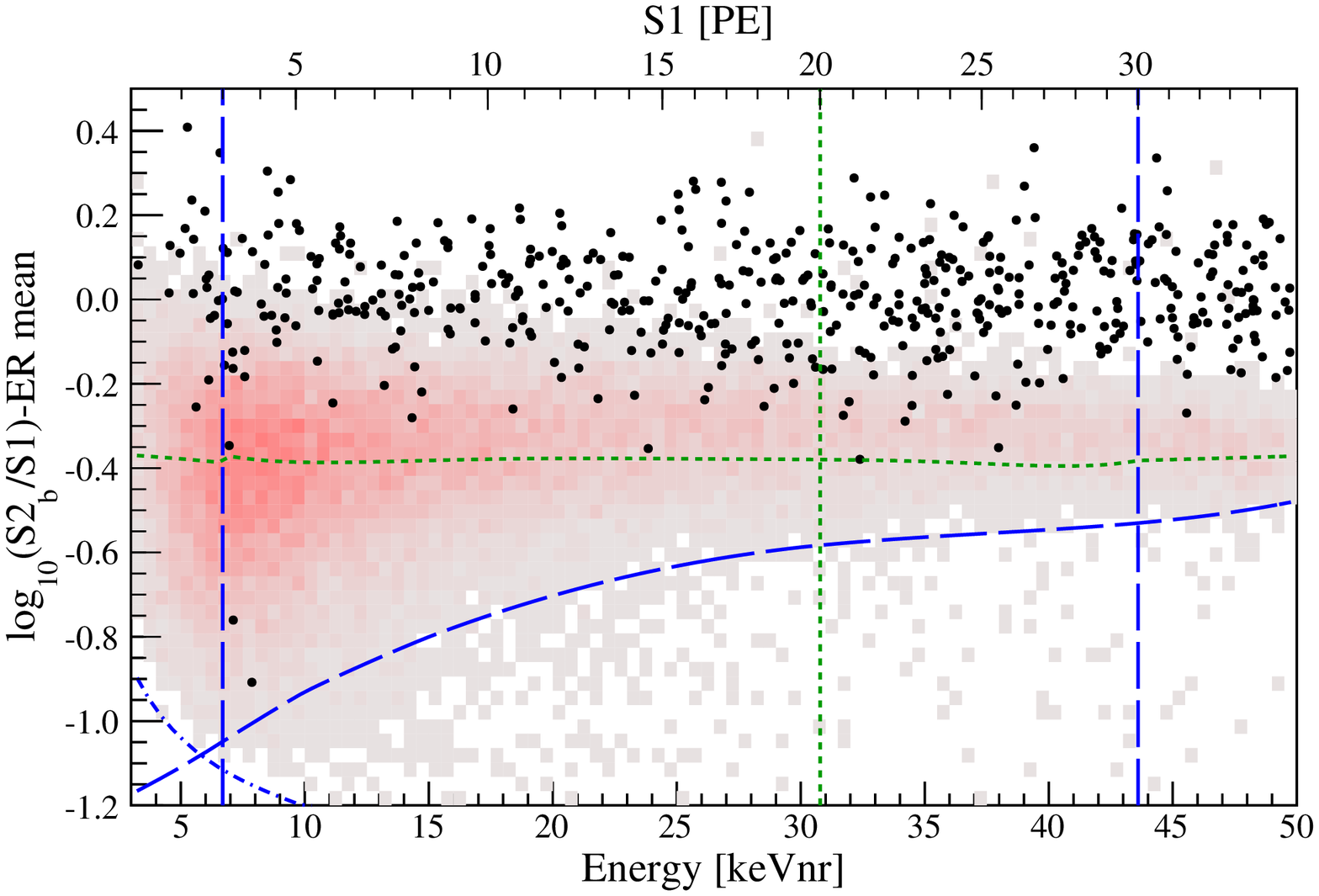}
\includegraphics[width=0.45\textwidth]{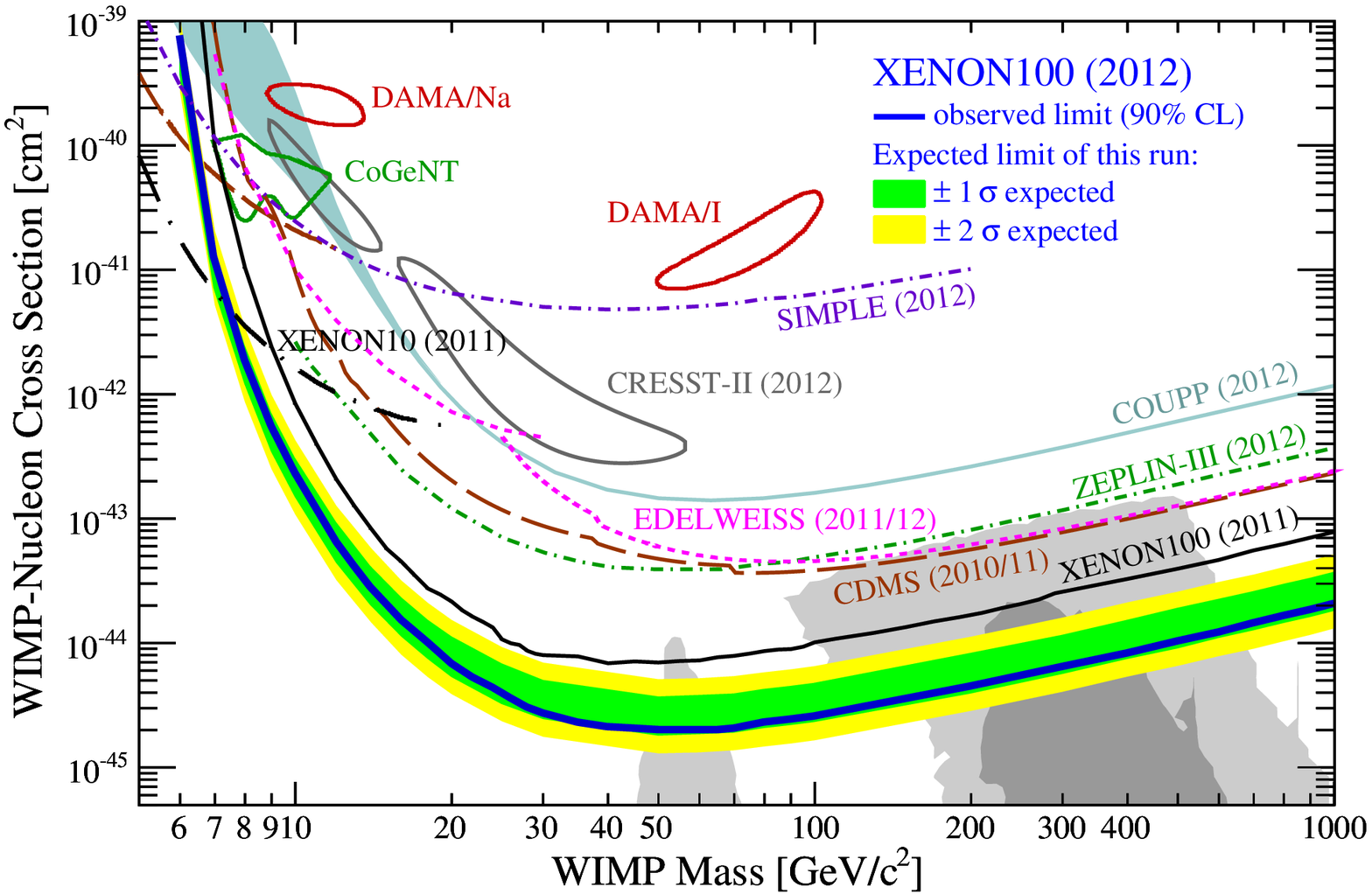}
\end{center}
\end{figure}

As in the previous run, it was decided a priori to use a Profile Likelihood (PL) statistical inference method to set WIMP limits~\cite{xenon100PL}. The WIMP search is restricted to the energy window 3-30PE, which in nuclear recoil equivalent energy corresponds to 6.6-43.3 keVnr. The blinding procedure consisted of masking the dark matter data from 2-100PE in S1 by keeping only the upper 90\% of the ER band, hiding more than 90\% of the signal region. After a unblinding, two events were observed in a predefined benchmark WIMP search region (with shorter energy range 6.6-30.5 keVnr, where 1.0$\pm$0.2 events were expected), as shown in figure \ref{fig:xenon100run10result} (left). In this figure, we can also observe the negligible impact of the S2$>$150PE threshold cut which is indicated by the dashed-dotted blue curve. The signal region is also restricted by a lower border running along the 97\% NR quantile (dashed blue curve).
%The waveforms for both events are of high quality and their S2/S1 value is at the lower edge of the NR band from neutron calibration. The PL analysis yields a p-value of $\ge$ 5\% for all WIMP masses for the background-only hypothesis indicating that there is no excess due to a dark matter signal.
The PL analysis yields a p-value of $\ge$ 5\% for all WIMP masses for the background-only hypothesis indicating that there is no excess due to a dark matter signal.

90\% confidence level exclusion limit for the spin-independent WIMP-nucleon cross section $\sigma_\chi$ is calculated, under standard assumptions of the Dark Matter halo~\cite{xenon100run10}. The expected sensitivity of this dataset in the absence of any signal is shown by the green (1$\sigma$) and yellow (2$\sigma$) bands in Figure \ref{fig:xenon100run10result} (right). The new limit is represented by the thick blue line. It excludes a large fraction of previously unexplored parameter space, including regions preferred by scans of the constrained supersymmetric parameter space. The new XENON100 data provide the most stringent limit for m$_\chi >$ 8 GeV/c$^2$ with a minimum of $\sigma_\chi$ = $2.0~\times~10^{-45}~cm^2$ at m$_\chi$ = 55 GeV/c$^2$. The new XENON100 result continues to challenge the interpretation of the DAMA~\cite{dama}, CoGeNT~\cite{cogent}, and CRESST-II~\cite{cresst2} results as being due to scalar WIMP-nucleon interactions.

\section{The XENON1T experiment}\label{subsec:xenon1t}

In parallel to the successful operations of XENON100, the Collaboration has already designed the next generation detector, XENON1T, with a fiducial mass of about 1 ton and a total mass of 2.4 tons. The goal is to increase the fiducial volume by a factor 10 and reduce the background by a factor 100. By doing so, the sensitivity for 50 GeV WIMPs will go down to $10^{-47}$ cm$^2$, deeply exploring the region allowed by MSSM models. At the time of writing, detector and infrastructures are 90\% fully designed, including the muon veto, the water shield, the cryostat, the TPC, the cryogenic system and the purification circuit which was a distillation column.

Scaling up the detector by a factor 10 also requires a substantial increase in the size of the infrastructures needed to store and handle the xenon. One key aspect is how to store the xenon when it is not inside the TPC. For this purpose, the Collaboration has designed a special storage system, called ReStoX ~\cite{restox} (Recovery and Storage of Xenon), consisting to a cryogenic tank able to keep about 3.6 tons of xenon, either in the liquid phase (by cooling and recirculating it through a purification system) or as a gas (withstanding a pressure of about 65 bar).

The XENON Collaboration has mastered the technology of liquid xenon TPC to a level that a detector at the ton scale is now possible. While the XENON100 detector is still providing the best results in the search for WIMP-nucleon scattering, the Collaboration is already actively working for the design and the construction of XENON1T.

\end{document}